\documentclass{iau} 
\usepackage{graphicx}

\title[JD 11.~~Insights on the stellar progenitors of black holes] 
{Stellar progenitors of black holes: insights from optical and infrared observations}

\author[I.F. Mirabel]   
{I.F. Mirabel$^{1,2}$}

\affiliation{$^1$Institute of Astronomy and Space Physics. CONICET - Universidad de Buenos Aires,   
Ciudad Universitaria, Av. Cantilo S/N, 1428 Buenos Aires - Argentina  \\ email: {\tt mirabel@iafe.uba.ar} 
\\[\affilskip]$^2$Laboratoire AIM-Paris-Saclay, CEA/DSM/Irfu−CNRS, CEA-Saclay, pt courrier 131, 91191 Gif-sur-Yvette, France \\email: {\tt felix.mirabel@cea.fr}}

\pubyear{2017}
\volume{324} 
\jname{New Frontiers in Black Hole Astrophysics}
\editors{Andreja Gomboc, \& Carole Mundell, eds.}
\begin{document}

\maketitle

\begin{abstract} Here are reviewed the insights from observations at optical and infrared wavelengths for low mass limits above which stars do not seem to end as luminous supernovae. These insights are: (1) the absence in archived images of nearby galaxies of stellar progenitors of core-collapse supernovae above 16-18 M$_{\odot}$, (2) the identification of luminous-massive stars that quietly disappear without optically bright supernovae, (3) the absence in the nebular spectra of supernovae of type II-P of the nucleosynthetic products expected from progenitors above 20 M$_{\odot}$, (4) the absence in color magnitude diagrams of stars in the environment of historic core-collapse supernovae of stars with $\geq$20 M$_{\odot}$. From the results in these different areas of observational astrophysics, and the recently confirmed dependence of black hole formation on metallicity and redshift of progenitors, it is concluded that a large fraction of massive stellar binaries in the universe end as binary black holes.  

\end{abstract}

\keywords{black hole physics, gravitational waves, X-rays: binaries, supernovae: general}

\firstsection 
\section{Introduction}

The question on the range of masses of stars that may collapse to form black holes (BHs) without exploding as energetic supernovae (SNe) is of topical interest for the incipient Gravitational-Wave Astrophysics. Strong natal SN kicks disrupt massive binary systems and eject black holes from their site of birth, reducing the numbers of binary black holes (BBHs), which may be prolific sources of gravitational waves. 

Most models on the evolution of massive stars predict that above $\sim$16-18 M$_{\odot}$ a large fraction of stars end as BHs by direct (\cite[Sukhbold et al. 2016]{Sukhbold_etal16} and reference therein), or failed supernova collapse (\cite[Lovergrove \& Woosley 2013)]{LovergroveWoosley 2013 and references therein}. In the following are reviewed the current results from observations at optical and infrared wavelengths that are used to contrast these theoretical predictions.

\section{Missing progenitors of $\geq$18 M$_{\odot}$ among core-collapse supernovae}

\begin{figure}[h]
\begin{center}
 \includegraphics[width=5.3in]{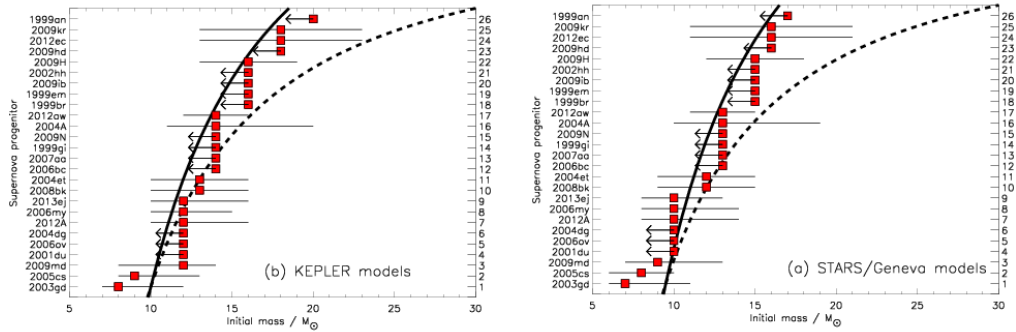} 
 \caption{Mass of stellar progenitors of core collapse SNe in the context of the STARS, Geneva, and KEPLER models of stellar evolution. The detections are marked with error bars, the limits with arrows, and the lines extend from the minimum to the maximum masses from cumulative Salpeter IMFs. Allowing the mass function to vary up to 30 M$_{\odot}$, the mass distribution need to be truncated at masses of $\sim$16.5 M$_{\odot}$ and $\sim$18M$_{\odot}$, respectively.  From Smartt (2015).}

   \label{fig1}
\end{center}
\end{figure}

Direct identifications of the progenitor stars of core-collapsed SNe discovered in nearby galaxies are being carried out by means of high resolution archival images from space and ground based telescopes. So far, among the 45 supernovae with either detected progenitors or upper limits, there has been a remarkable deficit of stars above an apparent limit of log L/L$_{\odot}$$\sim$5.1 dex (Smartt 2015). Type II-P SNe are the explosions of red super-giant stars (RSGs) which are believed to have masses in the range of 10 M$_{\odot}$ to 40 M$_{\odot}$. Figure \ref{fig1} illustrates the absence of luminous RSGs progenitors with estimated masses $\geq$ 16-18 M$_{\odot}$, from which it has been suggested (Smartt 2015) that the bulk of stars of $\geq$ 16-18 M$_{\odot}$ end as BHs with no visible supernova. Furthermore, with the lack of detected high mass progenitors in type IIb SNe, it has been proposed that “the missing high mass problem” has become relevant to all type II SNe (Smartt, 2015).

\section{Disappearance of massive stars without bright supernovae}

\begin{figure}[h]
\begin{center}
 \includegraphics[width=2.0in]{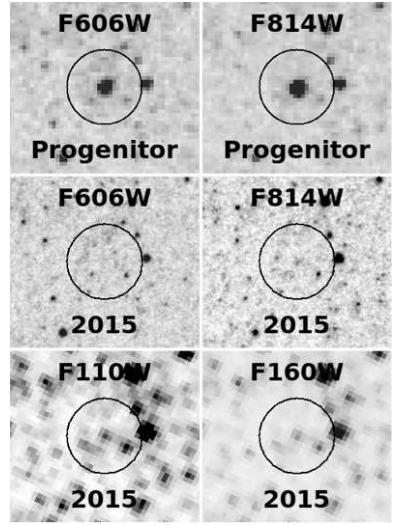} 
 \caption{HST images of the region surrounding N6946-BH1.
The top row shows the progenitor images with the filters WFPC2 F606W (588.7 nm) on the left, and F814 (802.4 nm) on the right, respectively. The middle row shows the corresponding 2015
WFC3 images. The bottom row shows the images with the WFC3/IR F110W (1153.4 nm) filter on the left, 
and F160W (1536.9 nm) on the right. The circles have a radius of 1 arcsec. The
progenitor has dramatically faded in the optical but in 2015 there is still
faint near-IR emission. From Adams et al. 2016.}

   \label{fig2}
\end{center}
\end{figure}

\begin{figure}[h]
\begin{center}
 \includegraphics[width=5.3in]{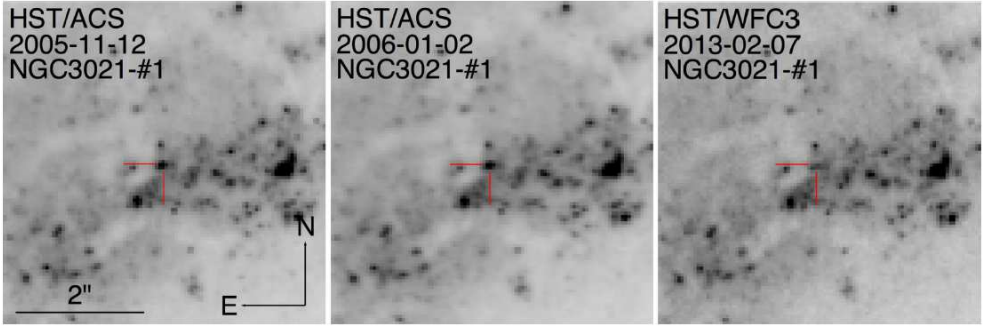} 
 \caption{Selected HST F814W image cuts centered on the position of a candidate in the galaxy NGC3021 with location indicated with tick marks. From Reynolds et al. (2015)}

   \label{fig3}
\end{center}
\end{figure}

An alternative approach is the search for massive stars that quietly disappear without bright supernovae by means of repeated observations of nearby galaxies (\cite[Kochanek et al. 2008]{Kochanek_etal08}). Following this strategy, from a survey of 27 galaxies with the Large Binocular Telescope (\cite[Gerke et al. 2015]{Gerke_etal15}), and a systematic analysis of archival Hubble Space Telescope images of 15 galaxies (\cite[Reynolds et al. 2015]{Reynolds_etal15}), were found respectively, one candidate now named N6946-BH1 with an estimated mass of $\sim$25 M$_{\odot}$, likely associated with a failed SN (Figure \ref{fig2}), and one candidate of 25-30 M$_{\odot}$ named NGC3021-1 that underwent an optically dark core-collapse (Figure \ref{fig3}). The optical disappearance of the $\sim$25 M$_{\odot}$ RSG first identified in the survey with the Large Binocular Telescope, now named N6946-BH1, has recently been confirmed using new and archival Hubble Space Telescope imaging  (\cite[Adams et al. 2016]{Adams_etal16}). 

The light curves of both candidates showed long time lasting faint transient displays before disappearing, as predicted in the model of very faint SNe from neutrino mass loss (\cite[[Lovergrove \& Woosley (2013)]{[LovergroveWoosley 2013}). From N6946-BH1 remains faint near infrared emission, and the mid-infrared emission measured with Spitzer slowly decreased to the lowest levels since the first measurements. The authors propose that this event is due to the ejection of the loose envelope of the red RSG that collapsed as a failed SN with a late-time emission due to fall-back accretion onto a newly formed black hole. Given the detection of 6 successful SNe in the sample of 27 monitored galaxies and one likely failed SN, the implied fraction of core-collapses that result in failed SNe is f$\sim$0.14$\pm$0.20 at 90$\%$ confidence, and if the current candidate is ultimately rejected, there is a 90$\%$  confidence upper limit on failed SN fraction of f$\leq$0.35.

\vskip.4in

\section{Indirect strategies for the mass limits of ccSN stellar progenitors} 

An indirect strategy was applied (\cite[Williams et al. 2014]{Williams_etal14}) using resolved stellar photometry from archival Hubble Space Telescope imaging to generate color magnitude diagrams of stars within 50 pc of the location of 17 historic core-collapse supernovae that took place in galaxies within a distance of 8 Mpc. Fitting the color magnitude distributions with stellar evolution models to determine the best-fit age distribution of the young population, the authors conclude that so far there is not a single high-precision measurement of a SN progenitor of $\geq$20 M$_{\odot}$. 

Furthermore, the nucleosynthetic products in the nebular spectra of supernovae can also provide constraints on the mass of the exploding star. In particular, it has been realized (\cite[Jerkstrand et al. 2014]{Jerkstrand_etal14}) that the observed evolution of the cooling lines of oxygen are difficult to reconcile with the expected nucleosynthesis products from progenitors of Type II-P SN with $\geq$20 M$_{\odot}$.

{\textbf{Caveats:} It has been argued that the optical and infrared observational approaches described above, and the conclusions inferred from them, may have several biases, such as those possibly due to the influence of circumstellar dust, magnitude variations of RSGs, luminosity-mass analysis, sample selection, and limited numbers statistics. Although those caveats could apply to particular cases, it is unlikely that such biases could account for the remarkable general stellar mass limit of 18-20 M$_{\odot}$ for the progenitors of Type II SNe found by the different observational methods at optical and infrared wavelengths presented here, and at radio wavelengths presented by Mirabel (2017) at another contribution of these proceedings. However, there may be mass islands of progenitor explodability as proposed by current models (e.g. \cite[Sukhbold et al. 2016]{Sukhbold_etal16}) that can produce successful supernovae for different ranges of progenitor masses. 

\section{Metallicity and redshift dependence of black hole formation} 

The metallicity (Z) dependence of BH formation has been observationally confirmed using a large set of extended off center x-ray luminosities (mostly coming from accreting BHs in high mass x-ray binaries) in nearby galaxies with known Zs and star formation rates (SFRs). For a given SFR the off-nuclear x-ray luminosities are typically ten times in low-Z galaxies ($\leq$20$\%$ solar) than in Z$_{\odot}$ galaxies (\cite[Douna et al. 2015]{Douna_etal15}). Besides, the X-ray luminosity due to accreting stellar BHs in normal galaxies show out to z=2.5 a redshift (z) an evolution given by 
L$_{2-10 keV}$/(HMXB)/SFR $\alpha$ (1 + z) due to declining Z with increasing z (\cite[Lehmer et al. 2016]{Lehmer etal16}).

\section{Conclusion}

•	A large fraction of stars above 18-20 M$_{\odot}$ may collapse by implosion and most likely end as BHs. This is inferred from searchers for SN progenitors in archived images at optical and infrared wavelengths, from massive stars that quietly disappear in the dark, from the largest stellar masses in the young stellar populations that host historic SN remnants, and from the absence of nucleosynthetic products of very massive stars in the nebular spectra of core-collapse SNe. 

•	The detection of an X-ray source at the position of a disappearing luminous star as proposed by Adams et al. (2016) may solve the problem of the missing high-mass SN progenitors, and open the possibility for the observation of BH formation in real time. Because most massive stars belong to multiple stellar systems, black holes recently formed in situ may accrete from nearby stars and produce X-rays and synchrotron radio waves, as BHs in high mass X-ray binaries. To witness the birth and early infancy of BHs, the sensitivities and angular resolutions of the James Webb Space Telescope for observations at 10-20 microns, Athena for observations at X-rays, and SKA for the observations at radio wavelengths may be required.  

•	From the metallicity dependence of BH formation and dynamical evolution of massive stellar systems, can be inferred to first order approximation that $\sim$30$\%$ of stellar binaries with Z$\leq$0.1 Z$_{\odot}$ and primaries above 40 M$_{\odot}$ end as binary black holes.

\end{document}